\documentclass[pra,superscriptaddress,longbibliography,twocolumn]{revtex4-1}
\usepackage[utf8]{inputenc}
\usepackage[english]{babel}
\usepackage{amsmath}
\usepackage{amsfonts}
\usepackage{amssymb}
\usepackage{graphicx}
\usepackage{amsthm}
\usepackage{color}
\usepackage{hyperref}
\usepackage[caption=false]{subfig}
\usepackage{graphicx,float}
\usepackage{dcolumn}
\usepackage{bm}
\usepackage{mathrsfs}
\usepackage{txfonts}
\usepackage{CJK}
\usepackage[amssymb]{SIunits}
\usepackage{epsfig}
\usepackage{epstopdf}
\usepackage{lipsum}
\usepackage{placeins}
\usepackage{physics}

\newenvironment{SChinese}{%
	\CJKfamily{gbsn}%
	\CJKtilde
	\CJKnospace}{}
\allowdisplaybreaks[2]

\begin{document}
	
	\begin{CJK}{UTF8}{}
		\begin{SChinese}
			
			\title{Reversible optical isolators and quasi-circulators using a
			magneto-optical Fabry-P\'{e}rot cavity}
			\author{Tiantian Zhang}  %
			\affiliation{College of Engineering and Applied Sciences, National 
			Laboratory of Solid State Microstructures, and  Collaborative 
			Innovation Center 
			of Advanced Microstructures, Nanjing University, Nanjing 210093, 
			China}
			
			\author{Wenpeng Zhou}  %
			
			\affiliation{College of Engineering and Applied Sciences, National 
				Laboratory of Solid State Microstructures, and Collaborative 
				Innovation Center of Advanced Microstructures, Nanjing 
				University, 
				Nanjing 210093, China}
			
			\author{Zhixiang Li}  %
			
			\affiliation{College of Engineering and Applied Sciences, National 
				Laboratory of Solid State Microstructures, and Collaborative 
				Innovation Center of Advanced Microstructures, Nanjing 
				University, 
				Nanjing 210093, China}
			
			\author{Yutao Tang}  %
			
			\affiliation{Shenzhen Shaanxi Coal Hi-tech Research Institute Co., 
			Ltd, Shenzhen 518083, China}
			
			\author{Fan Xu}  %
			
			\affiliation{Shenzhen Shaanxi Coal Hi-tech Research Institute Co., 
			Ltd, Shenzhen 518083, China}
			
			\author{Haodong Wu}  %
			
			\affiliation{College of Engineering and Applied Sciences, National 
				Laboratory of Solid State Microstructures, and Collaborative 
				Innovation Center of Advanced Microstructures, Nanjing 
				University, 
				Nanjing 210093, China}
			
			\author{Han Zhang}  %
			\affiliation{School of Physics, Nanjing University, Nanjing 210023, 
			China}
			
			\author{Jiang-Shan Tang(唐江山)}  %
			\email{js.tang@nju.edu.cn}
			\affiliation{College of Engineering and Applied Sciences, National 
			Laboratory of Solid State Microstructures, and  Collaborative 
			Innovation Center 
			of Advanced Microstructures, Nanjing University, Nanjing 
			210093, China}
		
			\author{Ya-Ping Ruan(阮亚平)}  %
			\email{ruanyaping@nju.edu.cn}
			\affiliation{College of Engineering and Applied Sciences, National 
				Laboratory of Solid State Microstructures, and  Collaborative 
				Innovation Center 
				of Advanced Microstructures, Nanjing University, Nanjing 
				210093, China}
			
			\author{Keyu Xia (夏可宇)}  %
			\email{keyu.xia@nju.edu.cn}
			\affiliation{College of Engineering and Applied Sciences, National 
			Laboratory of Solid State Microstructures, and Collaborative 
			Innovation Center of Advanced Microstructures, Nanjing University, 
			Nanjing 210093, China}
			\affiliation{Jiangsu Key Laboratory of Artificial Functional 
			Materials, Nanjing University, Nanjing 210023, China}

			
			\begin{abstract}
				Nonreciprocal optical devices are essential for laser 
				protection, modern optical communication and quantum 
				information processing by 
				enforcing one-way light propagation. The conventional Faraday 
				magneto-optical 
				nonreciprocal devices rely on a strong magnetic field, which is 
				provided by a 
				permanent magnet. As a result, the isolation direction of such 
				devices is fixed 
				and severely restricts their applications in quantum networks. 
				In this work, we 
				experimentally demonstrate the simultaneous one-way 
				transmission and 
				unidirectional reflection by using a magneto-optical 
				Fabry-P\'{e}rot cavity and 
				a magnetic field strength of $50~\milli\tesla$. An optical 
				isolator and a 
				three-port quasi-circulator are realized based on this 
				nonreciprocal cavity 
				system. The isolator achieves an isolation ratio of up to 
				$22~\deci\bel$ and an 
				averaged
				insertion loss down to $0.97~\deci\bel$. The quasi-circulator 
				is realized with 
				a 
				fidelity exceeding $99\%$ and an overall survival probability 
				of 
				$89.9\%$, corresponding to an insertion loss of $\sim 
				0.46~\deci\bel$. The  
				magnetic field is provided by an electromagnetic coil, thereby 
				allowing for reversing the light circulating path.  The 
				reversible 
				quasi-circulator paves the way for building reconfigurable 
				quantum networks.
			\end{abstract}
			
			\maketitle
			
		\end{SChinese}
	\end{CJK}
	
	\section{Introduction}\label{sec:intro}
	Nonreciprocal optical devices (NRODs), including optical isolators and
	circulators, are critical components in the classical optics regime and
	photonic quantum systems,\cite{Sathyamoorthy2014,Daiss2021,tang2022quantum} 
	as 
	they protect lasers and sensitive signals by isolating or separating the
	backscattered light.\cite{Caloz2018} In the quantum domain, circulators 
	play a 
	crucial role in the field of quantum information and are essential in 
	quantum 
	network architectures.\cite{science.354.2016,sliwa2015reconfigurable}
	
	The optical nonreciprocity that breaks the time-reversal symmetry of
	light propagation is usually attained by polarizers and Faraday
	rotators, which rely on magnetically biased
	materials.\cite{Linkhart2014,Jalas2013,Potton2004,Kamal2011}
	However, limited by the weak magneto-optical (MO) effect, such conventional
	NRODs typically require strong magnetic 
	fields,\cite{Gauthier1986,Ballato1995,Snetkov2014,Starobor2019} which are 
	generated by permanent magnets, hindering the reconfigurability of NRODs in 
	practical optical applications. Dynamically reversing the propagation 
	direction 
	of light in
	NRODs is highly desirable, particularly for reconfigurable 
	quantum networks. 
	\cite{Daiss2021,kang2011reconfigurable,huang2017dynamically,PhysRevLett.78.3221.1997,RevModPhys.87.1379.2015,nature.541.2017,natphotonics.12.744.2018,
		shen2018reconfigurable,ren2022nonreciprocal,kimble2008quantum,
		alshowkan2021reconfigurable,ren2023nonreciprocal}
	
	
	To circumvent the severe constraints imposed by strong
	magnetic fields, one effort is devoted to magnet-free optical
	nonreciprocity, including chiral quantum optics
	systems,\cite{zhang2019chiral,PhysRevA.90.043802.2014,PhysRevX.5.041036.2015,science.354.2016,nature.541.2017,PhysRevA.99.043833.2019,PhysRevLett.128.203602.2022,LPR.17.2023}
	spatiotemporal modulation of the
	medium,\cite{PhysRevLett.110.093901.2013,PhysRevLett.110.223602.2013,natphotonics.11.774.2017}
	optical
	nonlinearity,\cite{cotrufo2023passive,yang2019realization,natphotonics.8.524.2014,PhysRevLett.117.123902.2016,PhysRevLett.118.033901.2017,natelectron.1.113.2018,prj.9.1218.2021,
		tang2022quantum, CPL.39.124201}
	the Doppler
	effect,\cite{natphotonics.12.744.2018,PhysRevLett.121.203602.2018,PhysRevResearch.2.033517.2020,PhysRevLett.125.123901.2020,sciadv.7.12.2021,aqt.5.8.2022}
	optomechanical
	resonators,\cite{oe.20.7.2012,natphys.11.3.2015,nphotonics.10.10.2016,Fang2017,nphotonics.15.1.2021}
	spinning resonators,\cite{PhysRevLett.121.153601.2018} etc. An
	alternative avenue involves enhancing the MO effect by exploiting strong MO
	materials,\cite{chai2020non,Fan2019,Carothers2022,toyoda2019nonreciprocal} 
	and cavity-enhanced
	strategies.\cite{	
		shen2018reconfigurable,ren2021non,nphoton.5.758.2011,PhysRevLett.123.023602.2019,optica.7.11.2020}
	Moreover, unidirectional invisibility 
	also attracts intensive attention, engineering light reflection in 
	unprecedented manners across diverse systems and 
	strutures.\cite{lin2011unidirectional,wang2019nonreciprocity,wu2014non,feng2013experimental,inoue2023unidirectional}
	The simultaneous one-way transmission and unidirectional reflection 
	is fundamentally interesting, but its realization remains a 
	challenge.
	
	Thanks to the fact that light repeatedly passes through the
	object in an optical resonant system, the interaction between light and
	matter can be greatly enhanced in a Fabry--P\'{e}rot (F-P)
	cavity.\cite{wallsquantum.2007} The use of  F-P cavities to amplify
	Faraday rotation for achieving isolator effects has been extensively
	studied in theory,\cite{Ling1994,Dong2010,Rosenberg1964,Li2006} but it has
	rarely been experimentally verified. This configuration is known as an
	MO F-P (MOFP) cavity\cite{Ling1994} or a resonant optical
	Faraday rotator.\cite{Rosenberg1964} Furthermore, the implementation of
	more feature-rich reversible optical circulators or quasi-circulators using
	this system is still elusive. 
	
	In this paper, we demonstrate reversible optical isolators and three-port
	quasi-circulators based on the MOFP cavity. In comparison to
	conventional  MO nonreciprocal devices, our scheme requires a smaller 
	magnetic 
	field. Notably, this approach enables electrically controlling the magnetic
	direction, thereby enabling reversible optical isolators and
	quasi-circulators. Moreover, we achieve the simultaneous one-way 
	transmission and unidirectional reflection.
	The reported reversible NRODs may pave the way for reconfigurable quantum 
	networks.\cite{Daiss2021, 
		kang2011reconfigurable,huang2017dynamically,PhysRevLett.78.3221.1997,RevModPhys.87.1379.2015,nature.541.2017,natphotonics.12.744.2018,shen2018reconfigurable,ren2022nonreciprocal,kimble2008quantum,alshowkan2021reconfigurable,ren2023nonreciprocal}
	
	
	\section{System and model}\label{sec:model}
	Our reversible optical isolator and quasi-circulator consist of an MOFP 
	cavity and two
	sets of polarization beam splitters (PBSs) and quarter wave plates (QWPs).
	Their schematic diagrams for opposite isolation directions are shown
	in Figs.~1(a) and 1(b). The MOFP cavity is composed of a F-P
	cavity and a piece of terbium gallium garnet (TGG) crystal, a type of MO 
	crystal. The F-P cavity 
	supports two
	degenerate circularly polarized optical modes, $\sigma^{+}$ and
	$\sigma^{-}$, with the same frequency $\omega_0$. Here, to avoid confusion,
	$\sigma^{+}$- and
	$\sigma^{-}$-polarized fields are defined by the rotating direction of the
	electric-field vectors with respect to the magnetic field
	direction.\cite{Hu2021, LPR.17.2023} This is different from the optical
	case, where polarization is defined as either left-handed or right-handed
	circular polarization by observing the rotation of the electric field
	vector in relation to the direction of light 
	propagation.\cite{lipson2010optical}
	Therefore, the $\sigma^{+}$ and $\sigma^{-}$ polarizations can be either
	left-handed or right-handed circular polarization as defined in optics.
	This is essentially because the refractive index response of an
	MO crystal to optical polarizations depends on the rotation of
	the electric-field vectors relative to its quantization axis ($z$-axis),
	which corresponds to the direction of the magnetic field in our system. In
	the case of $\sigma^{+}$- and $\sigma^{-}$-polarized fields, the refractive
	indices of the MO crystal are denoted as $n_{+}$ and $n_{-}$,
	respectively. This leads to the breaking of the originally degenerate
	resonant frequencies $\omega_0$ into $\omega_{+}$ and $\omega_{-}$. 
	\begin{figure*}
		\centering
		\includegraphics[width=17cm]{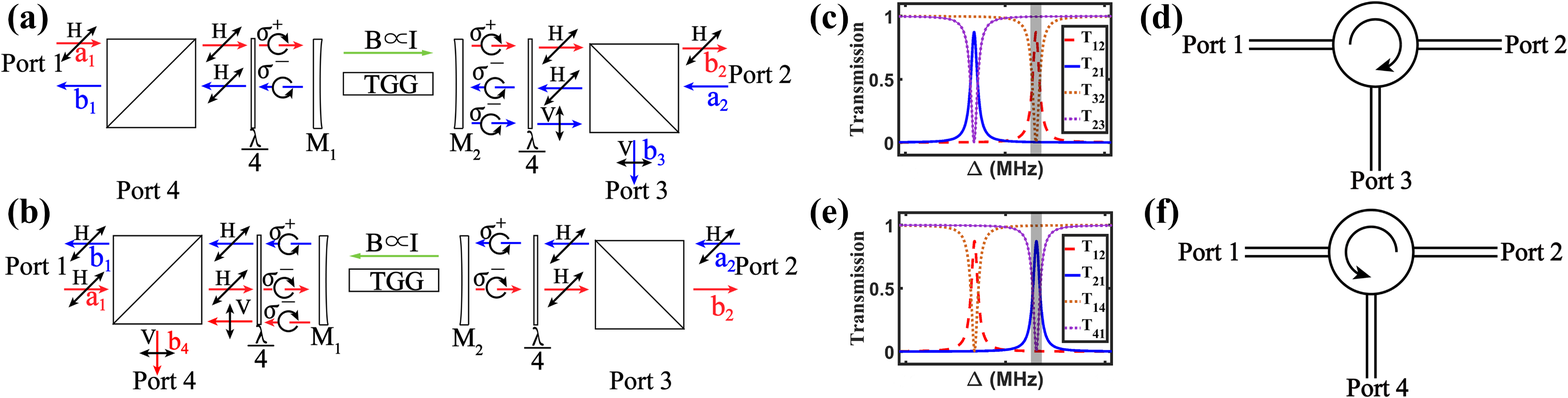}
		
		\vskip 2mm
		
		\caption{[(a), (b)] Conceptual schematics of the reversible 
			optical isolators and
			quasi-circulators. Experimental diagrams of the
			clockwise (CW) (a) and counterclockwise (CCW) (b) configurations 
			comprise a 
			magneto-optical Fabry--P\'{e}rot (MOFP) cavity, a pair of quarter 
			wave 
			plates
			(QWPs), and a pair of polarization beam splitters (PBSs), 
			respectively. 
			The
			magnetic direction of the CCW configuration is opposite to that
			of the CW configuration. The combination of PBSs and QWPs locks the 
			$\sigma^{\pm}$ polarization and the optical propagation direction 
			in 
			the 
			MOFP cavity. Black arrows indicate local optical polarization. 
			Here, 
			$\sigma^{+}$ and $\sigma^{-}$ polarizations are defined by the 
			rotating 
			direction of the electric-field vectors relative to the direction 
			of 
			the 
			magnetic field. The circulation of the polarized light is achieved 
			through 
			PBSs. TGG: terbium gallium garnet. [(c), (e)] Schematic 
			transmission spectra of the CW (c)
			and CCW (e) quasi-circulators versus the detuning 
			$\Delta = \omega -\omega_{0}$, respectively. The shaded areas
			indicate the operating band of the isolator and quasi-circulator. 
			[(d), 
			(f)] Schematic CW (d) and CCW (f) quasi-circulators depicting the 
			flowing 
			direction of light in the CW and CCW configurations,
			respectively.  }
	\end{figure*}
	
	{By utilizing two sets of the PBSs and QWPs, the locking of
		$\sigma^{\pm}$ polarization and the optical propagation direction can be
		realized in the MOFP cavity,\cite{Hu2021, LPR.17.2023} as shown by the
		black arrows in Figs.~1(a) and 1(b). In this system, the input
		field aligned with the magnetic direction exhibits $\sigma^{+}$
		polarization, whereas its antiparallel counterpart manifests as
		$\sigma^{-}$ polarization. This results in a chiral refractive index
		response within the MOFP cavity. In this case, the time-reversal 
		symmetry
		of the system is broken, leading to nonreciprocal resonant transmission
		spectra. However, in the absence of MO crystals, the system
		becomes reciprocal.}
	
	Next, we investigate the transmission properties of the 
	system, which can be expressed as:
	\cite{collett1984squeezing,gardiner1985input,wallsquantum.2007,combes2017slh}
	\begin{subequations}\label{eq:T R rate}
		\begin{align}
			&	T_{\pm} =
			\frac{\kappa_{\text{ex}}^2}{\delta_{\pm}^{2}+\kappa_{\pm}^{2}/4}\;,\\
			&	R_{\pm} =
			\frac{\delta_{\pm}^{2}+\kappa_{i}^{2}/4}{\delta_{\pm}^{2}+\kappa_{\pm}^{2}/4}\;
			,\
		\end{align}
	\end{subequations}
	where $T_{+} (T_{-})$ and $R_{+} (R_{-})$ denote the transmittance and
	reflectance of the $\sigma^{+}$($\sigma^{-}$) incident light, 
	respectively.The losses $\kappa_{\pm}=\kappa_{\text{ex, 
			1}}+\kappa_{\text{ex, 2}}+\kappa_{i}+\kappa_{i,\pm}$ correspond to 
			the 
	total
	decay rates of $\sigma^{+}$- and $\sigma^{-}$-polarized fields in the
	MOFP cavity. Here, $\kappa_{\text{ex, 1}} $ and 
	$\kappa_{\text{ex, 2}}$
	describe the extrinsic losses from the cavity mirrors, $\kappa_{i}$
	represents the intrinsic loss of the bare cavity, and $\kappa_{i,+}$
	($\kappa_{i,-}$) arises from the absorption of $\sigma^{+}$-
	($\sigma^{-}$-) polarized field by the MO crystal. We set 
	$\kappa_{\text{ex, 1}}=\kappa_{\text{ex, 2}} = \kappa_{\text{ex}}$ in the 
	theoretical model. The detunings is $\delta_{+}=\omega-\omega_{+}$
	($\delta_{-}=\omega-\omega_{-}$), where $\omega$ is the frequency of
	the input light. It is obvious that the system exhibits different
	transmission and reflection characteristics when different polarized
	light incidences occur. By the merit of this property, optical
	nonreciprocity can be realized.
	
	{Since the $\sigma^{+}$- and $\sigma^{-}$- polarizations are
		defined with respect to the magnetic field's direction, altering the
		orientation of the magnetic field allows for control over the 
		polarization
		of incident fields, as illustrated in Figs.~1(a) and 1(b). In
		the case of traditional Faraday MO devices, the requirement
		for strong magnetic fields presents a challenge when attempting to 
		change
		the magnetic field's direction. However, in our system, only a 
		relatively 
		weak magnetic field is necessary, thus eliminating this issue, which 
		will be
		thoroughly explained. Hence, our device
		exhibits reconfigurability through manipulation of the magnetic field
		orientation.
	}
	
	{We utilize the transfer matrix method to investigate the
		transmission properties of the system.\cite{PhysRevLett.125.013902.2020}
		The notations for field components $\{\bm{a}\}$ and $\{\bm{b}\}$ are 
		shown
		in Fig.~1. We first study the situation in
		Figure. 1(a). The transmission relation
		between the incident light vector
		$\bm{a}~(\bm{a}=\{{a_{1},a_{2},a_{3},a_{4}\}}^{T})$
		and the outgoing light vector
		$\bm{b}~(\bm{b}=\{b_{1},b_{2},b_{3},b_{4}\}^{T})$
		can be written as:
		\begin{equation}\label{forward matrix1}
			\left(\begin{array}{l}
				b_{2} \\
				b_{3}
			\end{array}\right) =\left(\begin{array}{cccc}
				T_{+} & 0 & R_{+} \\
				0 & R_{-} & 0
			\end{array}\right)\left(\begin{array}{l}
				a_{1} \\
				a_{2} \\
				a_{3}
			\end{array}\right) \;,
		\end{equation}
		where the subscript represents the incident and outgoing light of port 
		$i$.
		Fig.~1(c) shows the theoretical transmission spectra from
		ports 1 and 2, respectively. Consider the input within the shaded 
		frequency
		range in the Fig.~1(c). When incident light enters through
		port 1, it excites the $\sigma^{+}$- polarized mode. At the resonance, 
		that
		is, when $\delta_{+} = 0$, in the case of 
		$\kappa_{\text{ex}}\gg\kappa_{i} +
		\kappa_{i,\pm}$, $T_{+}\simeq1$ and $R_{+}\simeq0$ (see the dashed red
		curve), indicating that the forward light is nearly fully 
		transmitted with minimal reflection. However, when incident light 
		enters 
		through port 2, it corresponds to the excitation of the $\sigma^{-}$- 
		polarized field, resulting in
		$\delta_{-} \gg \kappa_{-}$ due to the different resonant frequencies. 
		This
		leads to $T_{-}\simeq0$ and $R_{-}\simeq1$ (see the solid blue curve), 
		signifying that the backward input light is nearly completely 
		reflected and transmits to port 3. Hence, the exotic propagation 
		characteristics of forward and backward light exhibit nonreciprocal 
		transmission and unidirectional 
		invisibility.\cite{lin2011unidirectional}
		By leading the reflected light to a port other than the input one, we 
		can 
		construct a quasi-circulator. The optical path implements a clockwise 
		(CW)
		three-port quasi-circulator, along port 1 $\rightarrow$ 2 $\rightarrow$
		3,\cite{tang2022quantum} as shown in Figs.~1(a) and (d).
		Synthesizing the above content, the 
		ideal transmission matrix of the CW quasi-circulator can be derived as:
		\begin{equation}
			T_\text{id, CW} = \left(\begin{array}{cccc}
				1 & 0 & 0 \\
				0 & 1 & 0
			\end{array}\right)\;.
		\end{equation}
		
		When the magnetic field is reversed, light propagates along opposite 
		direction, corresponding to the situation in
		Fig.~1(b). Similarly, the transfer matrix can be written as
		\begin{equation}\label{backward matrix1}
			\left(\begin{array}{l}
				b_{1} \\
				b_{4}
			\end{array}\right)=\left(\begin{array}{cccc}
				0 &T_{+} & R_{+} \\
				R_{-} & 0 & 0
			\end{array}\right)\left(\begin{array}{l}
				a_{1} \\
				a_{2} \\
				a_{4}
			\end{array}\right) \;.
		\end{equation}
		The reversal of the magnetic field causes the circularly polarized 
		fields
		excited by different ports to also reverse. Consequently, the light path
		exhibits a counterclockwise (CCW) circulation of photons (port 2
		$\rightarrow$
		1 $\rightarrow$ 4), see Figs.~1(e) and 1(f). Based on 
		this,  we obtain the ideal transmission matrix of the CCW 
		quasi-circulator, 
		\begin{equation}
			T_\text{id, CCW} = \left(\begin{array}{cccc}
				0 & 0 & 1 \\
				1 & 0 & 0
			\end{array}\right)\;.
		\end{equation}
		\hspace{-0.13cm}In the following, we refer to Figs.~1(a) and 1(b) as CW 
		and CCW configurations, 
		respectively.
		
		In this study, we designate 
		our three-port circulator as a "quasi-circulator". 
		In practical applications, this three-port quaisi-circulator can 
		fulfill the majority of requirements for light 
		circulation.\cite{tang2022quantum} 
		Furthermore, the system discussed above has the potential 
		to function as a four-port close-loop circulator, with 
		the 4$\times$4 transmission matrices. To illustrate the potential for 
		upgrading to a four-port circulator, we provide a detailed 
		schematic diagram and accompanying discussions in the Supplementary 
		Materials.
		
		We consider a MOFP cavity with an effective optical path $L$. When a 
		magnetic field $B$ is applied to the MO crystal with length $l$ and the 
		Verdet constant $V$, the optical rotation angle $\phi$ and the 
		resulting 
		shift are:\cite{Mikhaylovskiy2012}
		\begin{equation}\label{B impact}
			\left\{\begin{array} { l }
				{ \phi = V B l} \;, \\
				{ \omega_{+}-\omega_{-} = \dfrac{2\text{FSR}\times\phi 
				}{\pi}}\;,
			\end{array}\right.
		\end{equation}
		where FSR is the free spectral range (FSR) of the
		MOFP cavity. $\text{ FSR} = c/2L$, with $c$ being the speed of light 
		in vacuum, $c=3\times 10^8~\meter\per\second$. Thus, 
		the impact of the magnetic field on the frequency shift can
		be expressed as
		\begin{equation}\label{B impact2}
			\centering
			\omega_{+}-\omega_{-} = \frac{2\text{ FSR}\times V\times l\times 
			B}{\pi}
			\;.
		\end{equation}
		This equation shows that the relationship between frequency shift and 
		magnetic field
		strength is linear.
		
		Finally, we investigate the impact of magnetic field strength on the
		isolation
		ratio ($\mathcal{I}$).
		Based on Eqs.~\eqref{forward matrix1} and~\eqref{backward matrix1}, we 
		can
		derive an expression for the effect of magnetic field strength on the
		isolation ratio. Taking the CW quasi-circulator as an example and the 
		approximation $\kappa_{\text{ex}}\gg\kappa_{i}$, the
		isolation ratio of ports 1 and 2 is defined as
		\begin{equation}\label{eq:11}
			\begin{aligned}
				\mathcal{I} = &10\text{ log}_{10}(\frac{T_{12}}{T_{21}})\\
				=& 
				10\text{log}_{10}(\dfrac{T_{+}a_{1}+R_{+}a_{3}}{T_{-}a_{2}})\\
				\simeq&10\text{log}_{10}\left[\dfrac{4(\omega_{+}-\omega_{-})^{2}+\kappa_{-}^{2}}{\kappa_{+}^{2}}\right]
				\;.
			\end{aligned}	
		\end{equation}
		Here, we have assumed that
		$a_{1}=a_{2}=a_{3}$ to simplify the calculation.
		By substituting Eq.~\eqref{B impact2} into Eq.~\eqref{eq:11}, we further
		have
		\begin{equation}\label{eq:12}
			\begin{split}
				\centering
				\mathcal{I}	=
				&10\text{log}_{10}\left[\dfrac{16(\text{FSR}\times V\times 
				l\times
					B)^{2}+\pi^{2}\kappa_{-}^{2}}{\pi^{2}\kappa_{+}^{2}}\right]\;.
			\end{split}	
		\end{equation}
		Equation~\eqref{eq:12} reveals that the isolation ratio is expected to
		increase
		as the magnetic field strength increases. This also explains why 
		traditional
		Faraday MO devices require strong magnetic fields. However, in
		the MOFP cavity system, the existence of the cavity (allowing 
		$\text{FSR}$
		to be regulated) provides the possibility of achieving high isolation 
		ratio
		with a small magnetic field. The insertion loss (IL) of the isolator is 
		defined as:\cite{PhysRevLett.121.203602.2018,tang2022quantum}
		\begin{equation}\label{eq:13}
			\centering
			\text{IL} = -10\log_{10}(T_{f}) ,
		\end{equation}
		where $T_f$ represents the transparent direction. It is $T_{12}$ in 
		Fig. 
		1(a) or $T_{21}$ in Fig.1(b).
		
		The performance of a quasi-circulator can be
		quantified with the average photon survival
		probability $\eta$ and the fidelity $\mathcal{F}$. The 
		survival probability of the probe light is given by $\eta_{i} = 
		\sum_{k}T_{ik}$, and $T_{ik}$ represents the transmission rate of light 
		from port $i$ to port $k$.  
		The average photon survival probability for quansicirculator is $\eta =
		\sum_{i}\eta_{i}/2$.\cite{tang2022quantum,PhysRevLett.121.203602.2018}}
	The corresponding insertion loss of the quasi-circulator is $\text{IL} = 
	-10 
	\log_{10} 
	(\eta)$.
	The fidelity $\mathcal{F}$ is evaluated by examining
	the overlap between the renormalized transmission matrix $\tilde{T} =
	(T_{ij}\per\eta_{i})$ and the ideal one $T_{\text{id}}$. 
	The average operation fidelity of the
	quasi-circulator can be calculated 
	as:\cite{PhysRevLett.121.203602.2018,science.354.2016}
	\begin{equation}\label{fedylity}
		\centering
		\mathcal{F} =
		\frac{\mathrm{Tr}[\tilde{T}T_{\text{id}}^{T}]}{\mathrm{Tr}[T_{\text{id}}T_{\text{id}}^{T}]}
		\;.
	\end{equation}
	
	It is important to note that in this framework, $\eta$ and 
	$\mathcal{F}$  are not limited to describing quantum systems, and their 
	purpose is strictly characterize the performance of the quasi-circulator. 
	The significance of these symbols lies in the ability to quantify the level 
	of agreement between the actual scattering matrix and the ideal 
	scattering matrix of the quasi-circulator. We leverage the robust 
	mathematical frameworks and analytical methodologies established within the 
	field of quantum optics to better characterise our devices.  
	
	\section{Experimental Setup}\label{sec:FixedD}
	The schematic  experimental setup is illustrated in Fig.~2. 
	We experimentally implement the proposed scheme based
	on an MOFP cavity system embedded with a TGG crystal. The crystal is 
	$l=18~\meter\meter$ in length and coated with a broadband 
	transmission-enhancing 
	film on both facets. The coated film covers the entire operating frequency 
	band. The TGG crystal has an extremely high transmittance, $V= 
	78.5~\radian\per\tesla\per\meter$ and the refractive index $n_0 =1.95$ in 
	the absence of the magnetic field. 
	The MOFP cavity is
	constructed by using two high-reflection
	concave mirrors ($r=99\%$ reflectivity), denoted as $\text{M}_{1}$ and 
	$\text{M}_{2}$.
	Each mirror has a curvature radius of $100~\milli\meter$. The spatial 
	distance 
	between the two mirrors is about $182~\milli\meter$, yielding an effective
	optical length $L=200~\milli\meter$ for the whole MOFP cavity. We obtain 
	$\gamma = n_0 l/L=0.18$, representing the ratio between the effective 
	optical 
	paths of the crystal and the cavity.
	The MOFP cavity  
	is positioned between two sets of QWPs and PBSs, collectively forming the 
	optical momentum-spin locking apparatus.\cite{LPR.17.2023} 
	The frequency difference $\omega_{+}-\omega_{-}$ increases with the ratio 
	$\gamma$.\cite{zak1991magneto}
	
	\begin{figure}
		\centering
		\includegraphics[width=8cm]{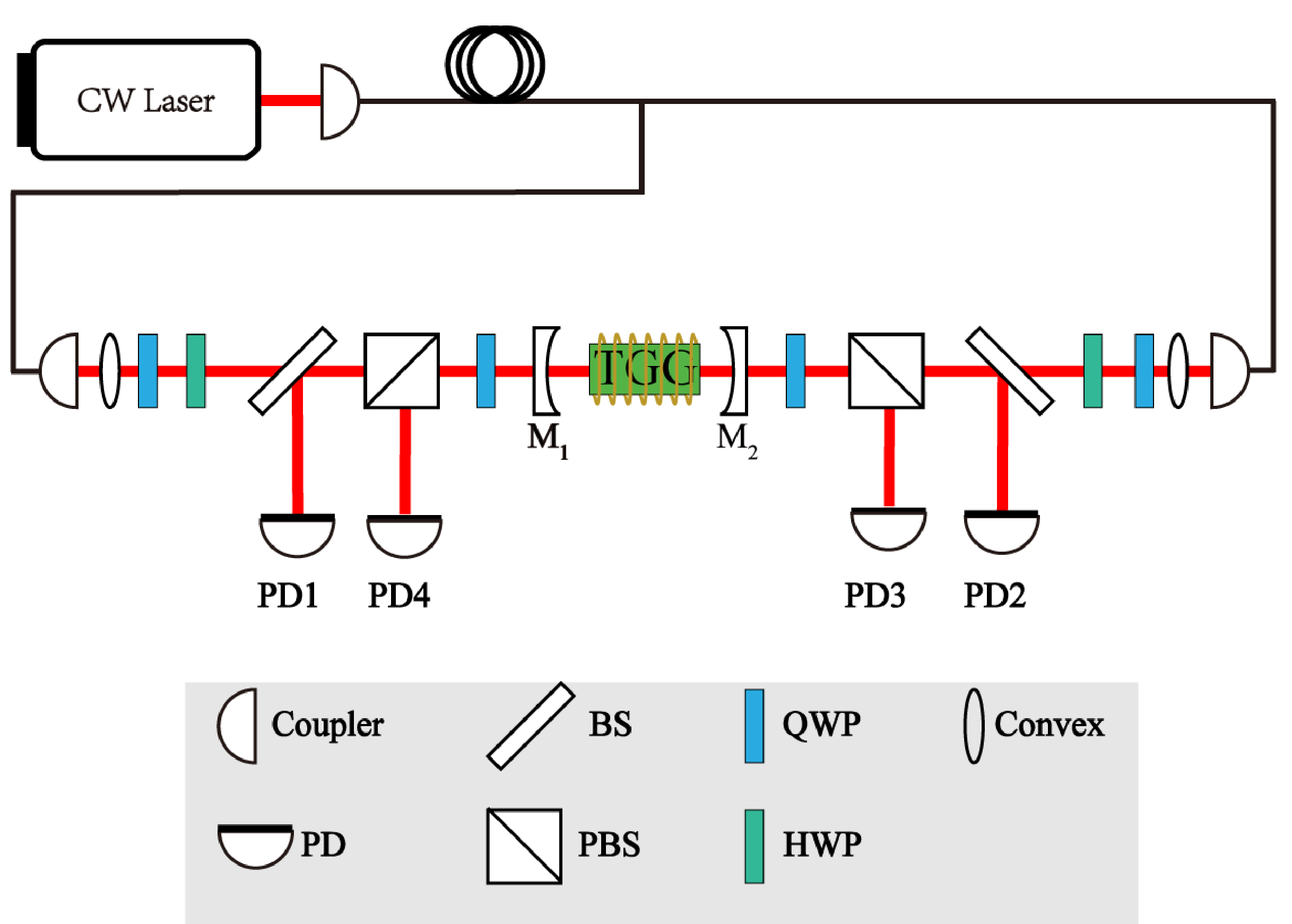}
		
		\caption{Schematic experimental setup for the optical isolators 
			and
			quasi-circulators based on the MOFP cavity. PD: photon detector; BS:
			beam splitter; PBS: polarization beam splitter; M: mirror; QWP:
			quarter-wave
			plate; HWP: half-wave plate; Convex: convex lens.}
	\end{figure}
	The fast axis of each QWP is oriented
	at an angle of $45~\degree$ relative to the horizontal direction, and the
	fast axes of the two QWPs are configured mutually orthogonally. Upon
	passing through the optical momentum-spin locking apparatus, all incident
	light undergoes spin modulation, resulting in distinct optical polarization
	states. Specifically, the light propagating within the MOFP cavity in the
	direction of the magnetic field is modulated into $\sigma^{+}$ polarized
	light, whereas the light traveling in the opposite direction is modulated
	into $\sigma^{-}$ polarized light. In addition, we insert a pair of BSs
	into the optical path so that the signal can be emitted from ports 1 and 2.
	A pair of PBSs enable the emission of signals with different polarizations
	from ports 3 and 4. In this way, the output beam of quasi-circulator can be
	monitored by the PD 1, PD 2, PD 3, and PD 4, which correspond to ports 1,
	2, 3, and 4, respectively. The experiment operates at 
	temperature of $25~\degree$C.
	
	The laser beams used in
	the experiment are derived from a tunable external cavity
	semiconductor laser operating at a wavelength of $795~\nano\meter$ and have
	a waist diameter of $159~\micro\meter$. These laser beams are directed
	towards the optical quasi-circulators via a fiber splitter
	that bifurcates the signal into two separate paths. Subsequently, they
	enter the optical quasi-circulator at ports 1 and port 2, respectively.
	
	An electromagnetic coil is used to generate a relatively weak magnetic 
	field for the MO crystal, as shown in Fig.~2. The relationship between the 
	strengths of the magnetic field ($B$) and the currents ($I$) is $B = \beta 
	I$ with the generating efficiency $\beta \approx 
	25~\milli\tesla\per\ampere$. It is worth noting that the magnetic field
	generated by the energized coil is significantly greater than that of the 
	Earth, thus making any potential impact from the latter negligible. 
	Therefore, 
	we can reverse the direction of the magnetic field by changing the 
	direction of
	the current. This feature allows for the straightforward reversal of the 
	optical
	isolators and quasi-circulators implemented within the system. 
	
	\section{Results}\label{sec:pi}
	To investigate the reversible optical nonreciprocal behavior, we
	measured the transmission of the beams in the system for both
	left-to-right (corresponding to CW configuration) and right-to-left
	(corresponding to CCW configuration) magnetic directions, as shown in
	Figs.~3 and 5, respectively. Swiftly switching
	between the two configurations can be achieved by changing the direction of
	the current. During our measurement, the $795~\nano\meter$ laser beams are
	incident simultaneously on both ports 1 and 2.

	\begin{figure}
		\centering
		\includegraphics[width=8cm]{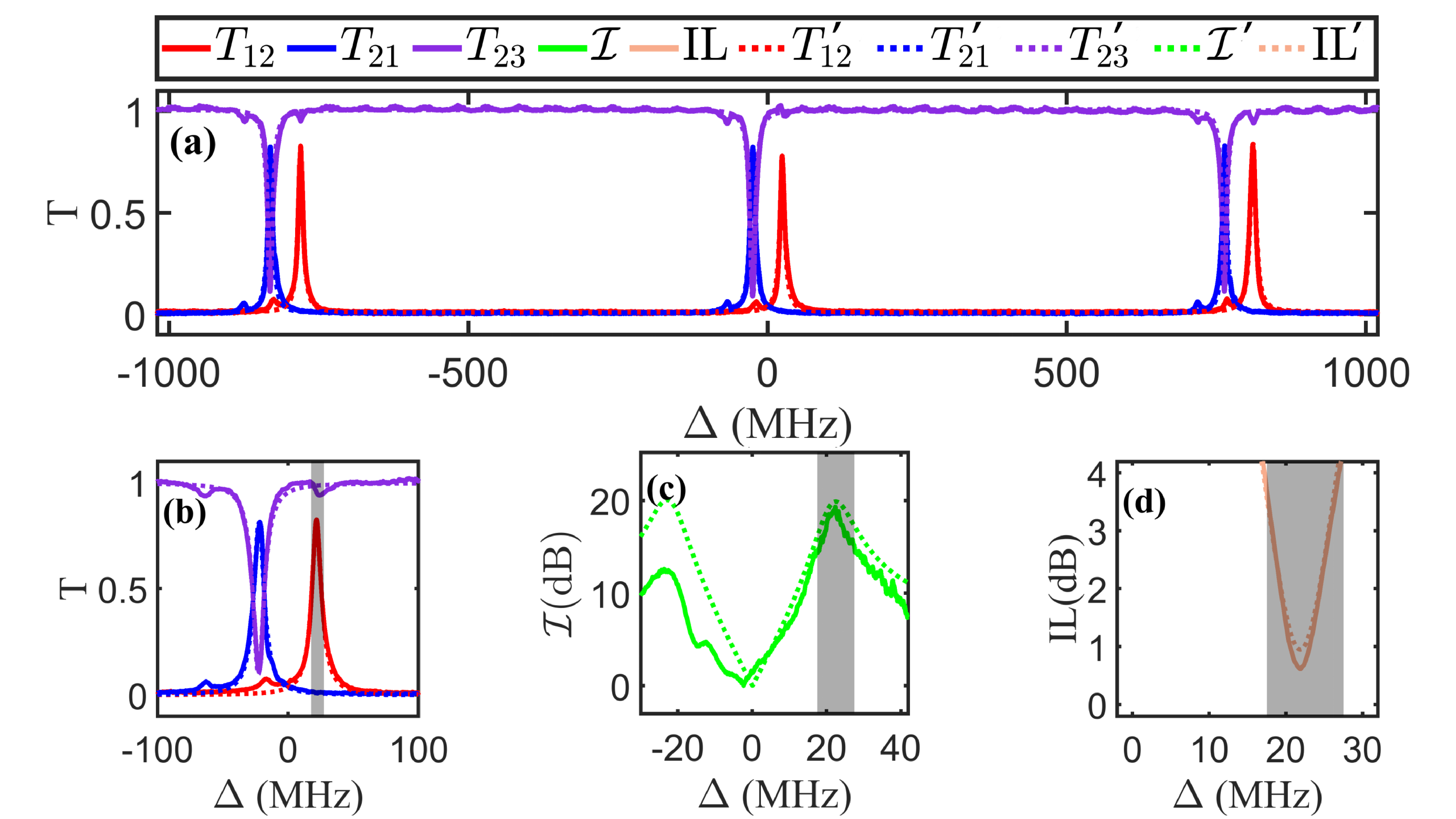}
		\vskip 2mm
		\caption{The performance of optical isolators based on CW
			configuration. [(a), (b)] The overall (a) and detailed (b) measured 
			transmission spectra between different ports. The solid curves 
			correspond 
			to the experimentally measured values, denoted as $T_{nm}^{'}$. The 
			dotted
			curves correspond to the theoretically
			calculated values, denoted as $T_{nm}$. In this context, $T_{nm}$
			signifies the transmission coefficient form port $m$ to port $n$,
			where $m, n \in  \{1, 2, 3, 4\}$. Fitting parameters: $\kappa_{ex} 
			= 
			4.3~\mega\hertz$, $\kappa_{i} =
			0.7~\mega\hertz$,
			$\kappa_{i,+} =
			0.16~\mega\hertz$, $\kappa_{i,-} = 0.4~\mega\hertz$, $B =
			50~\milli\tesla$.
			[(c), (d)]
			Isolation ratio (c) and insertion loss (d) of ports 1 and 2 as a 
			function
			of the detuning $\Delta$, respectively. The shaded regions indicate
			effective operating frequency range of the
			CW configuration.} \label{fig3}
	\end{figure}
	
	Initially, we characterize the optical nonreciprocity of the CW
	configuration.
	As demonstrated above, the magnetic field
	significantly enhances the circular dichroism of TGG crystal. The
	transmission spectra measured at each port of
	the CW configuration are shown in Fig.~3(a) for the analysis
	of multiple peaks and in Fig.~3(b) for the analysis of the
	individual peaks.

	It can be observed that the resonant frequency shifts as expected, thereby
	introducing nonreciprocal transmission within a specific frequency range
	(indicated by the shaded areas). When the incident light frequency matches
	the resonance frequency of the CW configuration, the transmission from port
	1 to port 2 is  $T_{12}(\omega_{+}) \approx 80\%$, corresponding to an 
	insertion 
	loss of $0.97~\deci\bel$, while
	the opposite transmission $T_{21}(\omega_{-}) \approx 0.45\%$. Thus, in CW 
	configuration,
	a high-performance optical isolator can be implemented with ports 1 and 2.
	The isolation ratio and insertion loss are shown in Figs.~3(c)
	and 3(d), respectively. The isolation ratio of incident beam at
	the resonance frequency($\omega_{+}$) reaches a maximum of
	$22.4~\deci\bel$.
	The resonance frequency splits by
	$\left| \omega_{+}-\omega_{-} \right| \approx \ 47.2~\mega\hertz$.

	Fig.~4 shows the isolation ratio versus the current strength. The green 
	curve 
	represents the experimental results for
	the isolation ratio, while the black curve is for the theoretical
	result obtained by substituting experimental parameters into
	Eq.~\eqref{eq:12}.
	When the current
	amplitude increases from $0~\ampere$ to $2~\ampere$, the effective magnetic 
	field strengths applied to the TGG crystal varies from
	$0~\milli\tesla$ to $50~\milli\tesla$. This magnetic field is considerably 
	less than that required for the conventional MO isolator. The latter 
	typically
	operates at the order of several Tesla
	magnitudes.\cite{OL11.82.1986,ncommun.4.1558.2013,LPL.17.015001.2019,OL.13.3471.2023}
	The isolation
	ratio increases as the magnetic field
	strength increases, with a maximum value greater than $20~\deci\bel$,
	implying that
	we can indeed control the isolation ratio of the isolator by adjusting the
	current intensity. In fact, the value of $\beta$ can be
	increased by the coil turn number, leading to a
	significant decrease in the required current.

	\begin{figure}
		\centering
		\includegraphics[width=7cm]{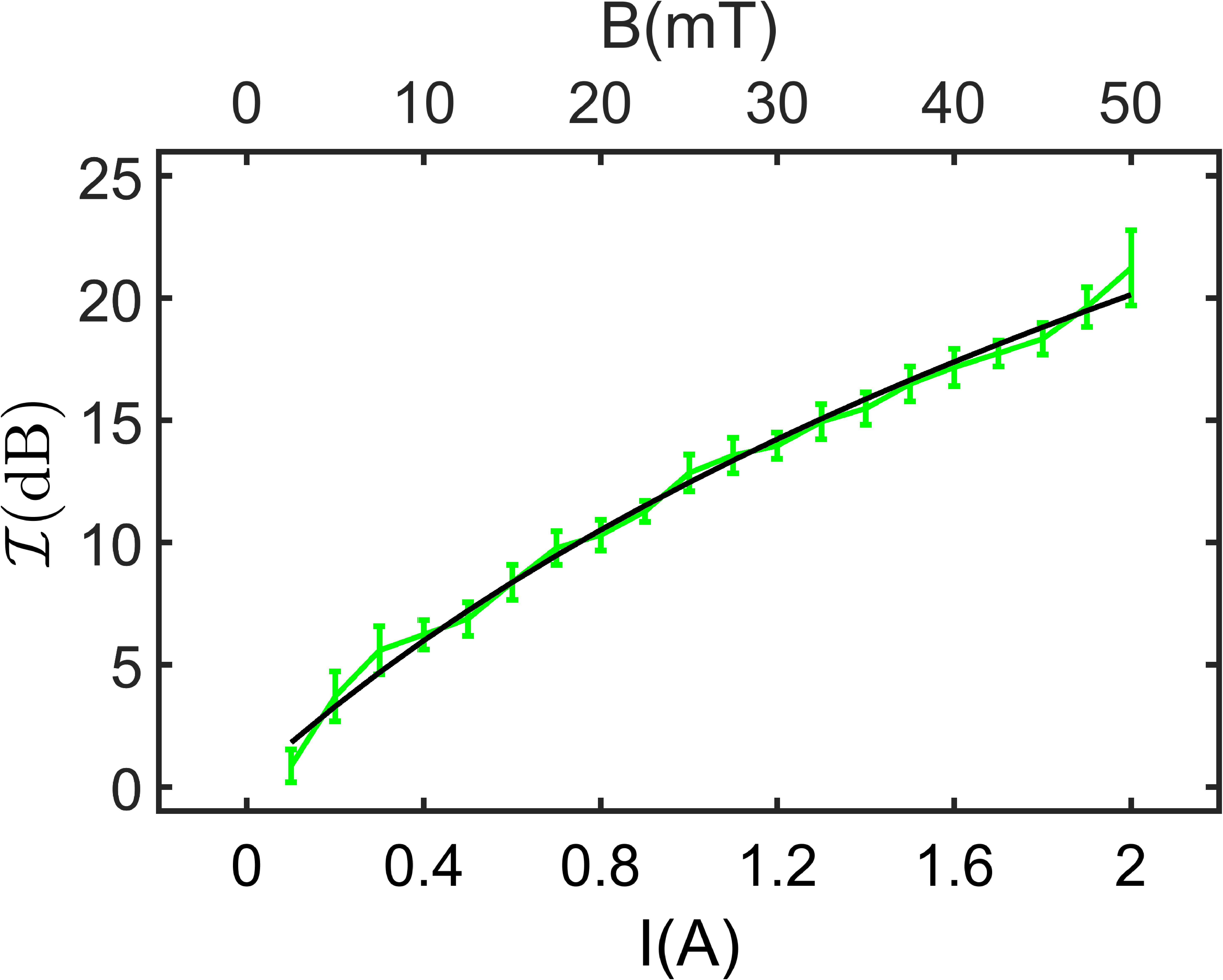}

		\caption{Isolation ratio in CW configuration versus the current
			strength and the magnetic field strength. Green curve calculated 
			from 
			experimental data; black curve is the theoretical results. Other 
			parameters 
			are as in Fig.~3.}
		
	\end{figure}

	For comparison with the CW configuration, the performance of the optical 
	isolator in the CCW
	configuration is also shown in Fig.~5. The magnetic field can be
	reversed by changing the direction of the current. In this case, when the 
	incident light
	frequency matches the resonance frequency of the CCW configuration without 
	the 
	need to adjust any other parameters of the system, see Figs. 
	5(a) and (b). The 
	transmission from port 2 to port 1 is $T_{21}(\omega_{+}) \approx 80.9\%$. 
	It is 
	much larger than the opposite transmission  $T_{12}(\omega_{-}) \approx 
	0.6\%$. 
	The corresponding insertion loss is $\text{IL} = 0.92~\deci\bel$. The
	frequency difference between the two resonance modes is
	$\left|\omega_{+}-\omega_{-}\right| \approx 47.1~\mega\hertz$. The 
	isolation 
	ratio and insertion loss of the CCW
	quasi-circulator are illustrated in Figs.~5(c) and 5(d),
	respectively. At the resonant frequency, the incident beam can achieve a
	maximum isolation ratio as high as $22.1~\deci\bel$. The bandwidth is about 
	$10~\mega\hertz$ for 
	$\mathcal{I}>11~\deci\bel$. Furthermore, all the theoretical results in
	Figs.~3-5 precisely match the corresponding
	experimental data. The CCW quasi-circulator shows the similar dependence of 
	the 
	isolation ratio on the current strength (not shown here).

	\begin{figure}
		\centering
		\includegraphics[width=8cm]{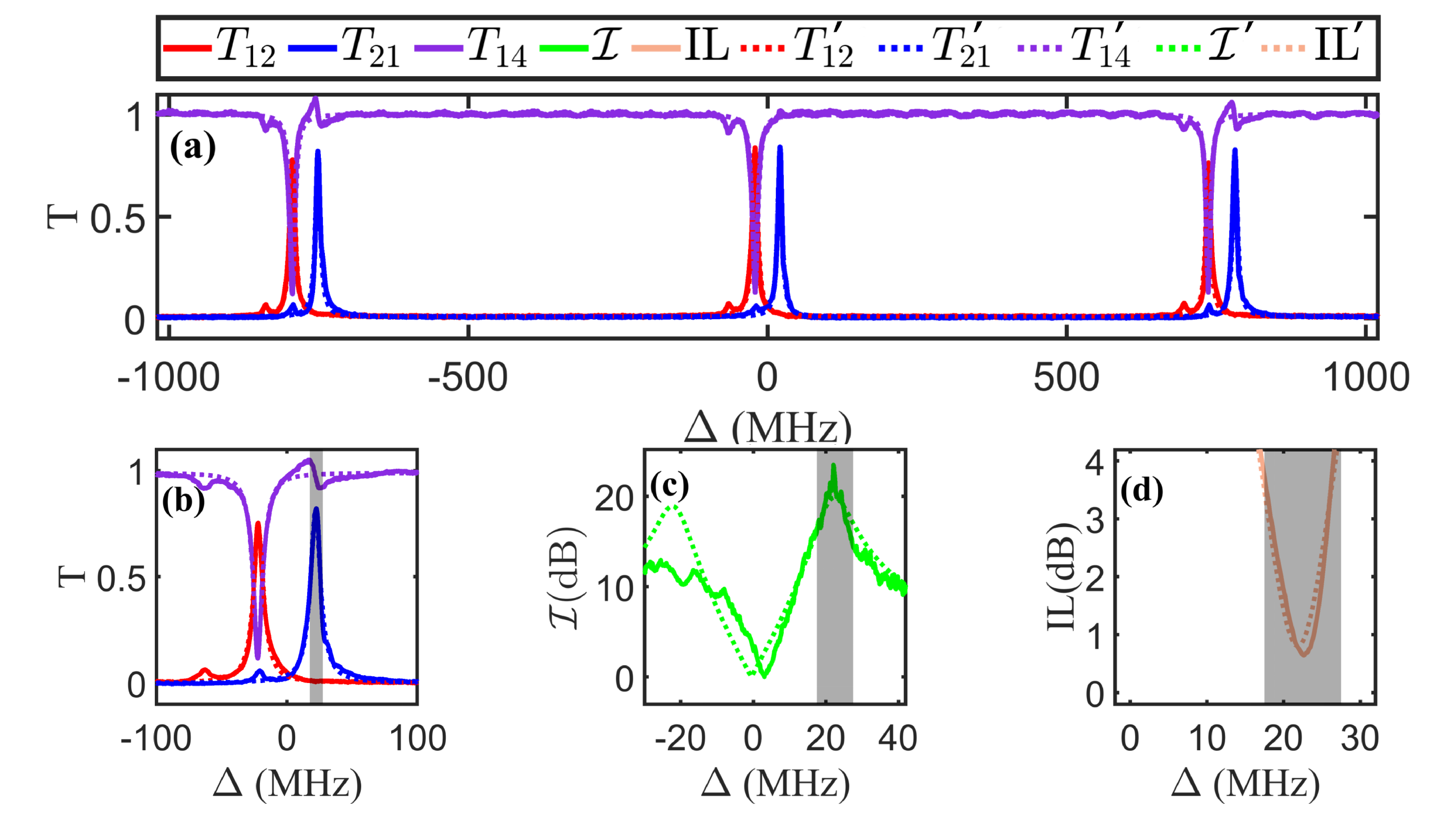}
		
		\caption{The performance of optical isolators based on CCW 
			configuration. [(a), (b)] The overall (a) and detailed (b) measured 
			transmission spectra between ports. The solid curves
			represent the experimentally measured values, labeled as 
			$T_{nm}^{'}$,
			while the dotted curves depict the theoretically calculated values 
			with the
			same parameters as in Fig.~3, denoted as $T_{nm}$. The fitting 
			parameters
			are $\kappa_{ex} = 4.3~\mega\hertz$, $\kappa_{i} =
			0.6~\mega\hertz$, $\kappa_{i,+} = 0.1
			~\mega\hertz$, $\kappa_{i,-} = ~0.4\mega\hertz$. [(c), (d)]
			Isolation ratio (c) and insertion loss (d) of ports 1 and 2 as a 
			function of
			the detuning $\Delta$, respectively. Similarly, the shaded regions
			indicate effective operating frequency range of the
			CCW configuration.}
		
	\end{figure}
	
	It is worth noting that we also observe unidirectional reflection in both 
	the 
	CW and CCW configurations. The reflected light can lead to a third port 
	to 
	form a quasi-circulator. Now we evaluate the performance of the three-port 
	quasi-circulators. In the CW configuration in Fig.~1(c), the transmission 
	$T_{23}$ from port 2
	to port 3 can reach about $99\%$ within the nonreciprocal region, see the 
	purple 
	curve in Fig. 3(b). Therefore, the system exhibits the
	function of an optical quasi-circulator with light flowing along
	port $1\rightarrow2\rightarrow3$. The ideal and measured transmission 
	matrices 
	of the CW 
	quasi-circulator are displayed in Figs.~6(a) and 6(c). The fidelity reaches 
	$\mathcal{F} \approx 99.5\%
	$ and the average photon survival probability is $\eta 
	= 89.4 \pm 0.4\%$.

	Similarly, by reversing the magnetic field, a CCW photonic quasi-cyclic
	direction from port 2 to port 1 and then to port 4
	(port $2\rightarrow1\rightarrow4$) can be established using the CCW
	configuration. We refer to this as a CCW three-port quasi-circulator.
	Ideal and measured transmission matrix are shown in Figs.~6(b) and 6(d). In 
	the 
	CCW three-port quasi-circulator, we can achieve a fidelity of $\mathcal{F} 
	\approx 99.6\% $ and an average photon survival probability of $\eta = 90.4 
	\pm 
	0.1\%$. The averaged survival probability of two cases is $89.9\%$, 
	yielding an 
	insertion loss of $\sim 0.46~\deci\bel$.

	\begin{figure}
		\centering
		\includegraphics[width=8cm]{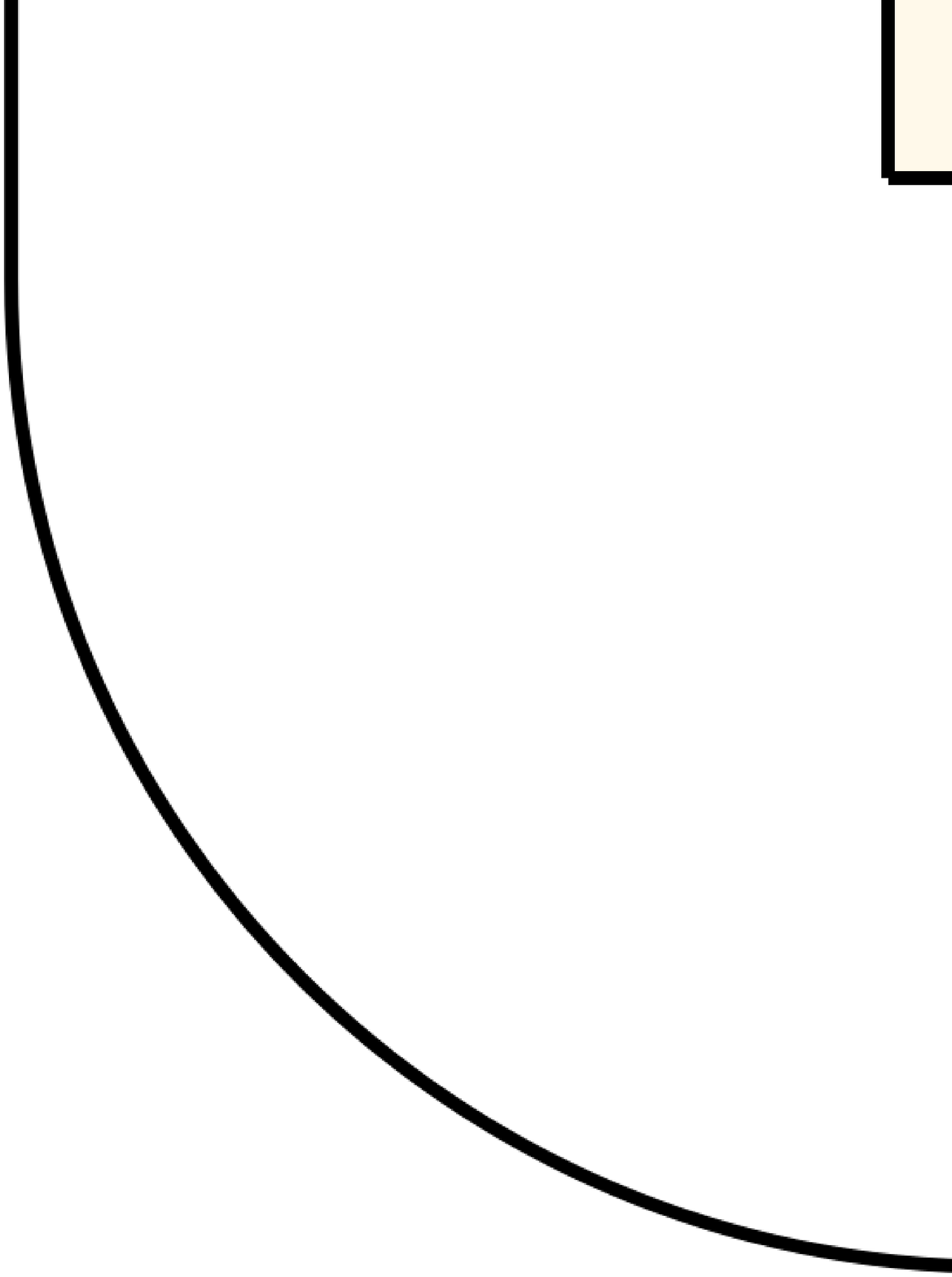}
		
		\vskip 2mm
		
		\caption{Transmission matrices of optical quasi-circulators. 
			[(a), (c)] Ideal (a) and measured (c) transmission 
			matrices of the CW quasi-circulator. [(b), (d)] Ideal (b) and 
			measured (d) 
			transmission matrices of the CCW quasi-circulator. Numbers inside 
			the 
			colore 
			squares represent the transmission between the two ports. Zero 
			transmissions 
			are indicated in the white regions. Other parameters are the same 
			as those in 
			Fig.~3.}
	\end{figure}
	
	\begin{figure}
		\centering
		\includegraphics[width=4.5cm]{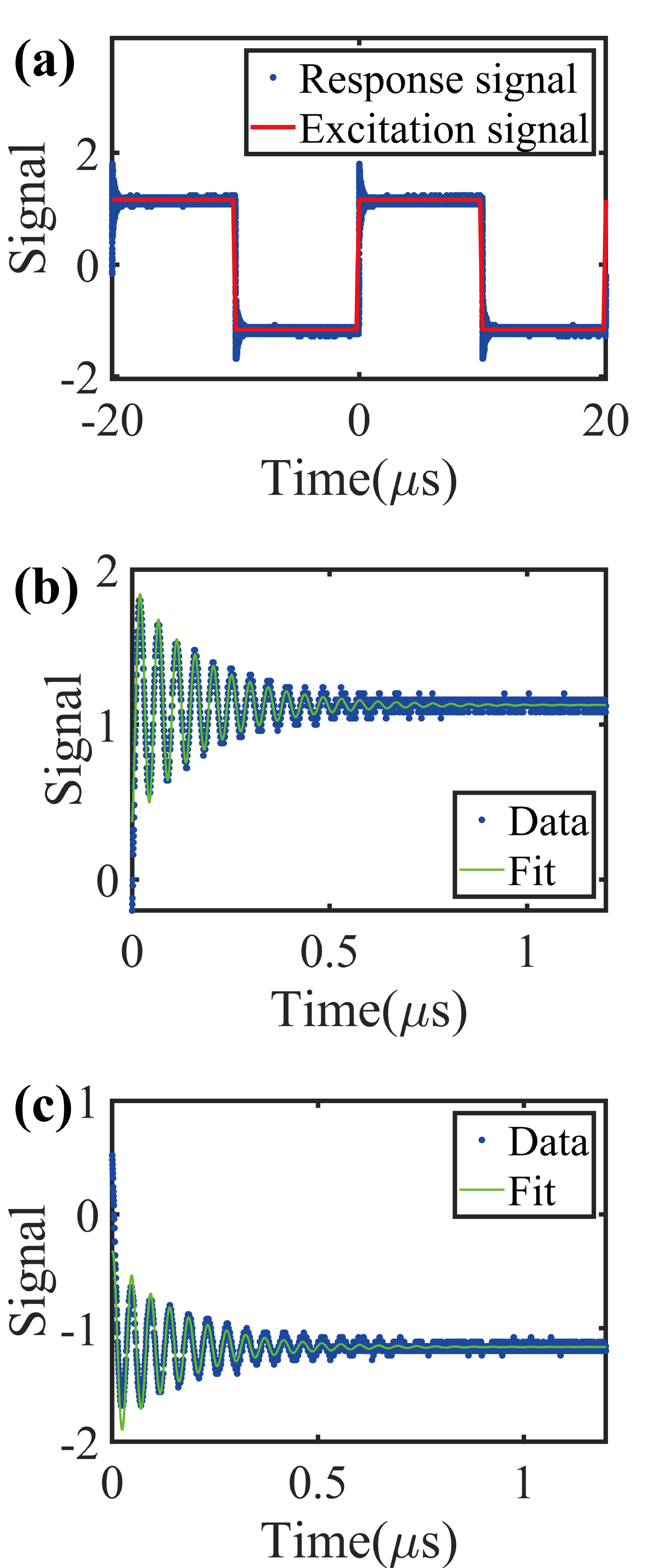}
		
		\vskip 2mm
		
		\caption{Performance of the system switching. (a)Overall performance 
			of the 
			system signals. The red line represents the waveform of the 
			excitation 
			signal generated by the power supply, while the blue  dots 
			represent 
			the 
			waveform of the response signal detected by the 
			oscilloscope. [(b), (c)]The waveform of the rising edge (b) and 
			falling 
			edge (c) of the actual detected response signal, where the blue  
			dots 
			represent the actual detected waveform, and the green curve 
			represents 
			the 
			theoretically fitted waveform.}
	\end{figure}
	
	It is worth mentioning that although we typically manually 
	change the direction of the magnetic field when measuring spectral line 
	data, 
	experimental results show that the system can switch between the CW and CCW 
	configurations in just $1~\micro\second$, as shown in Fig.~7. Furthermore, 
	through more sophisticated circuit design, this switching time can be 
	further 
	reduced, providing a very promising prospect for practical applications of 
	our system.

	\section{Conclusion}\label{sec:conc}
	In summary, we have demonstrated the reversible isolators and three-port 
	optical 
	quasi-circulators via combination of the MOFP cavity
	and special polarization modulation.
	The optical isolator achieves an isolation
	ratio of $22~\deci\bel$ and a low insertion loss of $0.97~\deci\bel$ under a
	$50~\milli\tesla$ magnetic field. This magnetic field is significantly 
	lower 
	than that required in commercial MO
	nonreciprocal devices and is provided by an electromagnetic 
	coil.\cite{OL11.82.1986,ncommun.4.1558.2013,Snetkov2014,LPL.17.015001.2019,OL.13.3471.2023}
	The fidelity of the three-port 
	quasi-circulator exceeds $99\%$ while maintaining 
	an overall averaged survival probability close to $89.9\%$, corresponding 
	to a 
	loss of about $0.46~\deci\bel$. Such low-loss reversible quasi-circulator 
	is 
	vital for reconfigurable quantum networks. 
	
	The magnetic field needed for our devices can be decreased 
	further by employing MO materials with higher Verdet
	constants, such as yttrium iron garnet (YIG), $V > 
	10^{3}~\rad\per\tesla\per\meter$.\cite{vojna2019verdet,ikesue2018development}
	On the other hand, larger spectral 
	splitting can be obtained by increasing the ratio 
	$\gamma$.\cite{zak1991magneto} When $\gamma$ increases to $0.9$, only a 
	$2~\milli\tesla$ magnetic field  is needed for the our quasi-circulators if 
	using the YIG crystal. Furthermore, in the case of 
	employing magnetic-optical materials with $V > 
	10^{5}~\rad\per\tesla\per\meter$,\cite{optica.7.11.2020,Fan2019} 
	the magnetic field required for our device would be weaker than 1 
	Gauss.
	
	\section*{Acknowledgements}
	This work was supported by National Key R\&D Program of China (Grant 
	No.~2019YFA0308700), the National Natural Science Foundation of China 
	(Grant Nos. 11890704, 92365107, and 12305020), the Program for Innovative 
	Talents and Teams in Jiangsu (Grant 
	No.~JSSCTD202138), the Shccig-Qinling Program, the China Postdoctoral 
	Science 
	Foundation (Grant No.~2023M731613), and the Jiangsu Funding Program for 
	Excellent Postdoctoral Talent (Grant No.~2023ZB708).

	%
	
	
	%
	
\end{document}